\documentclass[a4paper,11pt]{article}
\usepackage{jinstpub} 
\usepackage{lineno}
\usepackage{placeins}
\usepackage{float}
\usepackage{booktabs}
\usepackage{graphicx}

\title{Results from the GIF++ Long-Term Study of Eco-Friendly Gas Mixtures for RPCs}

\author[1]{D. Ahmadi}
\author[1,2]{M. Tytgat}
\author[4,5]{M. Abbrescia}
\author[21]{G. Aielli}
\author[5,10]{R. Aly}
\author[18]{M. C. Arena}
\author[12]{M. Barroso}
\author[6]{L. Benussi}
\author[6]{S. Bianco}
\author[9]{D. Boscherini}
\author[17]{F. Bordon}
\author[9]{A. Bruni}
\author[19]{S. Buontempo}
\author[17]{M. Busato}
\author[21]{P. Camarri}
\author[14]{R. Cardarelli}
\author[5]{L. Congedo}
\author[12]{D. De Jesus Damiao}
\author[5,8]{M. De Serio}
\author[5]{F. Debernardis}
\author[21]{A. Di Ciaccio}
\author[21]{L. Di Stante}
\author[15]{P. Dupieux}
\author[20]{J. Eysermans}
\author[4,5]{N. Ferrara}
\author[3,13]{A. Ferretti}
\author[5,8]{G. Galati}
\author[3,13]{M. Gagliardi}
\author[17]{R. Guida}
\author[4,5]{G. Iaselli}
\author[15]{B. Joly}
\author[24]{S.A. Juks}
\author[4,5]{U. Lakshmaiah}
\author[23]{K.S. Lee}
\author[14]{B. Liberti}
\author[22]{D. Lucero Ramirez}
\author[17]{B. Mandelli}
\author[15]{S.P. Manen}
\author[9]{L. Massa}
\author[5]{A. Pastore}
\author[14]{E. Pastori}
\author[6]{D. Piccolo}
\author[14]{L. Pizzimento}
\author[9]{A. Polini}
\author[14]{G. Proto}
\author[4,5]{G. Pugliese}
\author[3,13]{L. Quaglia}
\author[4,5]{D. Ramos}
\author[17]{G. Rigoletti}
\author[14]{A. Rocchi}
\author[9]{M. Romano}
\author[11]{P. Salvini}
\author[21]{R. Santonico}
\author[7]{G. Saviano}
\author[14]{M. Sessa}
\author[5,8]{S. Simone}
\author[3,13]{E. Vercellin}
\author[16]{M. Verzeroli}
\author[22]{N. Zaganidis}

\affiliation[1]{Inter-University Institute for High Energies, Vrije Universiteit Brussel, Pleinlaan 2, 1050 Brussels, Belgium}
\affiliation[2]{Ghent University, Dept. of Physics and Astronomy, Proeftuinstraat 86, 9000 Ghent, Belgium}
\affiliation[3]{INFN Sezione di Torino, Via P. Giuria 1, 10125 Torino, Italy}
\affiliation[4]{Politecnico di Bari, Dipartimento Interateneo di Fisica, via Amendola 173, 70125 Bari, Italy}
\affiliation[5]{INFN Sezione di Bari, Via E. Orabona 4, 70125 Bari, Italy}
\affiliation[6]{INFN - Laboratori Nazionali di Frascati, Via Enrico Fermi 54, 00044 Frascati (Roma), Italy}
\affiliation[7]{Sapienza Università di Roma, Dipartimento di Ingegneria Chimica Materiali Ambiente, Piazzale Aldo Moro 5, 00185 Roma, Italy}
\affiliation[8]{Università degli studi di Bari, Dipartimento Interateneo di Fisica, Via Amendola 173, 70125 Bari, Italy}
\affiliation[9]{INFN Sezione di Bologna, Via C. Berti Pichat 4/2, 40127 Bologna, Italy}
\affiliation[10]{Helwan University, Helwan, Cairo Governorate 4037120, Egypt}
\affiliation[11]{INFN Sezione di Pavia, Via A. Bassi 6, 27100 Pavia, Italy}
\affiliation[12]{Universidade do Estado do Rio de Janeiro, R. São Francisco Xavier 524, 20550-013 Maracanã, Rio de Janeiro - RJ, Brazil}
\affiliation[13]{Università degli studi di Torino, Dipartimento di Fisica, Via P. Giuria 1, 10125 Torino, Italy}
\affiliation[14]{INFN Sezione di Roma Tor Vergata, Via della Ricerca Scientifica 1, 00133 Roma, Italy}
\affiliation[15]{Clermont Université, Université Blaise Pascal, CNRS/IN2P3, Laboratoire de Physique Corpusculaire, BP 10448, F-63000 Clermont-Ferrand, France}
\affiliation[16]{Université Claude Bernard Lyon I, 43 Bd du 11 Novembre 1918, 69100 Villeurbanne, France}
\affiliation[17]{CERN, Espl. des Particules 1, 1211 Meyrin, Switzerland}
\affiliation[18]{Università degli studi di Pavia, Corso Strada Nuova 65, 27100 Pavia, Italy}
\affiliation[19]{INFN Sezione di Napoli, Complesso universitario di Monte S. Angelo ed. 6, Via Cintia, 80126 Napoli, Italy}
\affiliation[20]{Massachusetts Institute of Technology, 77 Massachusetts Ave, Cambridge, MA 02139, USA}
\affiliation[21]{Università degli studi di Roma Tor Vergata, Dipartimento di Fisica, Via della Ricerca Scientifica 1, 00133 Roma, Italy}
\affiliation[22]{Universidad Iberoamericana, Dept. de Fisica y Matematicas, 01210 Mexico City, Mexico}
\affiliation[23]{Korea University, 145 Anam-ro, Seongbuk-gu, Seoul, Korea}
\affiliation[24]{Université Paris-Saclay, 3 rue Joliot Curie, Bâtiment Breguet, 91190 Gif-sur-Yvette, France}

\emailAdd{donya.ahmadi@vub.be}

\abstract{Gaseous detectors play a crucial role in high-energy physics experiments, yet their operation often relies on fluorinated gases with a very high Global Warming Potential (GWP). Resistive Plate Chambers (RPCs) operated in avalanche mode typically use high-performance gas mixtures based on high-GWP F-gases such as C$_2$H$_2$F$_4$ and SF$_6$.
As environmental concerns grow and regulations become increasingly stringent, the development of sustainable gas mixtures has become a global priority.
Within the RPC ECOGas@GIF++ Collaboration, a long-term R\&D program has investigated eco-friendly gas mixtures for RPCs and assessed their performance under irradiation. RPCs operated with an HFO-1234ze/CO$_2$ mixture were exposed to high particle fluxes at CERN GIF++, accumulating $\mathcal{O}(10^2~\mathrm{mC/cm^2})$ over three years. This ageing campaign is now complete, and the detector performances have been systematically evaluated across a broad range of particle rates.
This paper presents preliminary final results from the campaign, along with future perspectives.}

\begin{document}
\maketitle
\flushbottom
\FloatBarrier
\section{Introduction}
\label{sec:intro}

Resistive Plate Chambers (RPCs) are gaseous detectors made of two parallel resistive
electrodes separated by a gas gap of order 1-2mm. Owing to their relatively low cost,
$\mathcal{O}(\mbox{ns})$ time resolution and $\mathcal{O}(\mbox{mm})$ spatial resolution, they are widely used
in high-energy physics for muon tracking, triggering and particle identification.
RPCs are usually operated in avalanche mode with a mixture of $\mathrm{C_2H_2F_4}$
($>90\%$), i-$\mathrm{C_4H_{10}}$ ($<5\%$) and $\mathrm{SF_6}$ ($<1\%$), where
$\mathrm{C_2H_2F_4}$ and $\mathrm{SF_6}$ are fluorinated greenhouse gases with a
Global Warming Potential (GWP) of $\approx 1430$ and $\approx 22800$, respectively.
Increasingly strict regulations~\cite{eu_2014_517, eu_2024_573} make the search for
sustainable alternatives a priority.
\begin{table}[!htb]
    \centering
    \caption{Composition and environmental impact of the gas mixtures.}
    \label{tab:gasmixtures}
    \resizebox{0.8\textwidth}{!}{
    \begin{tabular}{lccccccc}
        \hline
        & R134a & HFO-1234ze & CO$_2$ & i-C$_4$H$_{10}$ & SF$_6$ & GWP & CO$_2$e \\
        & (\%) & (\%) & (\%) & (\%) & (\%) & & (g/l) \\
        \hline

        STD  & 95.2 & -- & -- & 4.5 & 0.3 & 1485 & 6824 \\
        ECO2 & -- & 35 & 60 & 4 & 1 & 476 & 1522 \\
        ECO3 & -- & 25 & 69 & 5 & 1 & 527 & 1519 \\

        \hline

        Density (g/l)
        & 4.68 & 5.26 & 1.98 & 2.69 & 6.61 & -- & -- \\

        GWP
        & 1430 & 7 & 1 & 3 & 22800 & -- & -- \\

        \hline
    \end{tabular}
    }
\end{table}

Within the RPC ECOGas@GIF++ Collaboration, alternative, more eco-friendly mixtures based on tetrafluoropropene
(HFO-1234ze, $\mathrm{C_3H_2F_4}$) diluted with $\mathrm{CO_2}$ have been extensively
studied. The two mainly considered candidate mixtures, referred to as ECO2 and ECO3,
are compared with the standard mixture (STD) in table~\ref{tab:gasmixtures}.
A long-term ageing campaign was started at the CERN
Gamma Irradiation Facility (GIF++) to test the stability of RPCs operated with these
mixtures under sustained irradiation~\cite{ecogas_nima}.
This paper presents the preliminary results of the campaign and future
perspectives.

\FloatBarrier
\section{Experimental setup at GIF++}
\label{sec:setup}

\begin{figure}[H]
    \centering
    \includegraphics[width=0.5\textwidth]{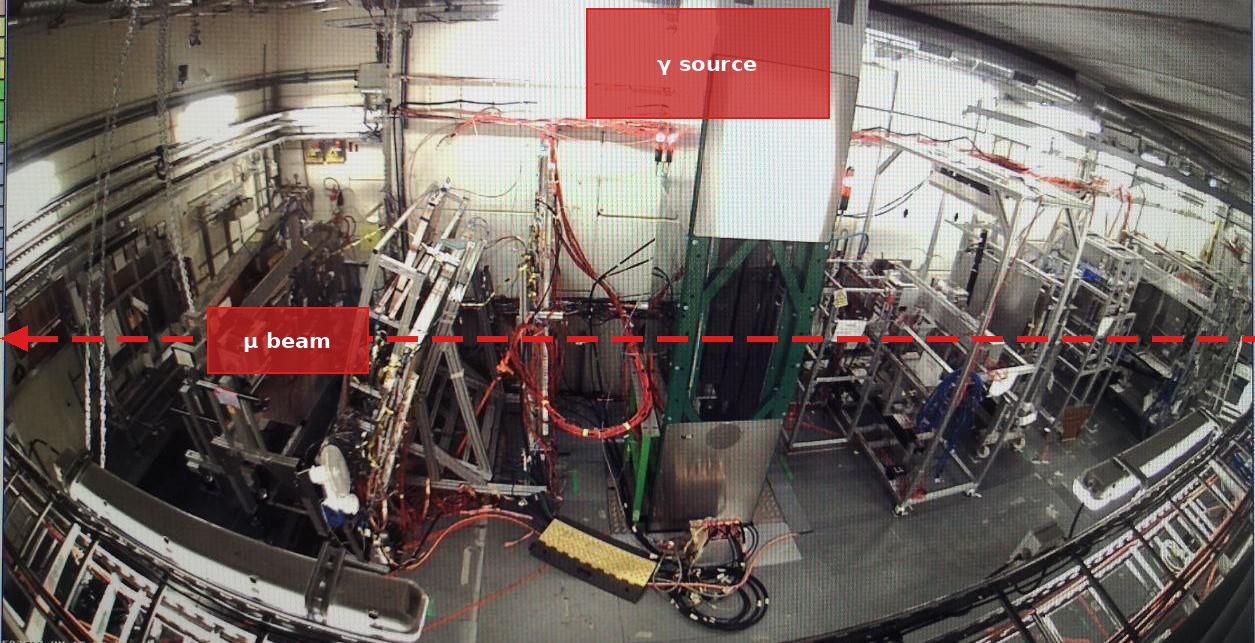}
    \caption{Layout of the experimental area at CERN GIF++.}
    \label{fig:gifpp}
\end{figure}

The layout of the experimental setup that the collaboration is operating at the CERN GIF++ facility is shown in Figure~\ref{fig:gifpp}. The facility is equipped with a
12.5~TBq $^{137}\mathrm{Cs}$ source that provides a high-radiation
environment~\cite{PFEIFFER201791}. Adjustable absorption filters
(ABS) installed in front of the source attenuate the radiation
field, enabling 27 different irradiation intensities. During dedicated
periods, a high-energy ($\sim 150$~GeV) muon beam from the secondary
SPS H4 beam line is also available at GIF++, and is used here to measure the RPCs' muon detection efficiency.

The detectors are mounted on two mechanical supports at GIF++ located approximately 3~m and 6~m away from the radiation source. During beam test campaigns, a scintillator-based trigger system is also installed. The RPCs are provided by various LHC-related experimental groups and differ in geometry, number and thickness of gas gaps, and readout electronics, as summarized in
Table~\ref{tab:detectors}.

\begin{table}[h]
\centering
\caption{Detectors installed at GIF++.}
\label{tab:detectors}
\begin{tabular}{lccr}
\toprule
RPC & Gap thickness & Electronics & Distance (m) \\
\midrule
ALICE       & 2 mm                & FEERIC + TDC  & 6   \\
ATLAS       & 2 mm                & Digitizer     & 3.2 \\
CMS RE1/1   & 2 mm -- double gap  & CMS FEB + TDC & 3.5 \\
CMS upgrade & 1.4 mm -- double gap & CMS FEB + TDC & 3.5 \\
EP-DT       & 2 mm                & Digitizer     & 3   \\
LHCb/SHiP  & 1.6 mm              & FEERIC + TDC  & 6.2 \\
\bottomrule
\end{tabular}
\end{table}

\FloatBarrier
\section{Long-term ageing studies}
\label{sec:ageing}
The ageing campaign operating the RPCs with the ECO2 gas mixture started in July 2022. All chambers were
flushed in open-loop mode with the ECO2 mixture and were operated at a fixed $HV_\mathrm{eff}$,
corrected for temperature and pressure. The dose rates to which chambers were exposed typically ranged from 1
to 5 mGy/h, depending on their position relative to the GIF++ source, corresponding to background rates due to the $\gamma$ irradiation between 400 and
1000 ~Hz/cm$^2$. The ageing study of RPCs operated with eco-friendly gas mixtures comprised three complementary activities: monitoring of the dark currents, measurement of the electrode resistivity, and evaluation of the RPC performance in beam tests after the irradiation campaign.

The high voltage (HV) and the currents drawn by each chamber were recorded every 30~s with the webDCS monitoring tool~\cite{Fagot2020}. Once a week, HV scans were performed with the $^{137}\mathrm{Cs}$ source fully shielded to measure dark current (the current without irradiation). Below the gas multiplication threshold ($\approx7$--$8$~kV for a 2~mm gap), part of this current flows through conductive paths in the chamber rather than through the gas, and therefore does not contribute to gas-induced ageing of detectors. This Ohmic component was obtained from a linear fit in the low-voltage region and subtracted from the current measured under irradiation to calculate the integrated charge.

\begin{figure}[htbp]
    \centering
    \includegraphics[width=0.57\textwidth]{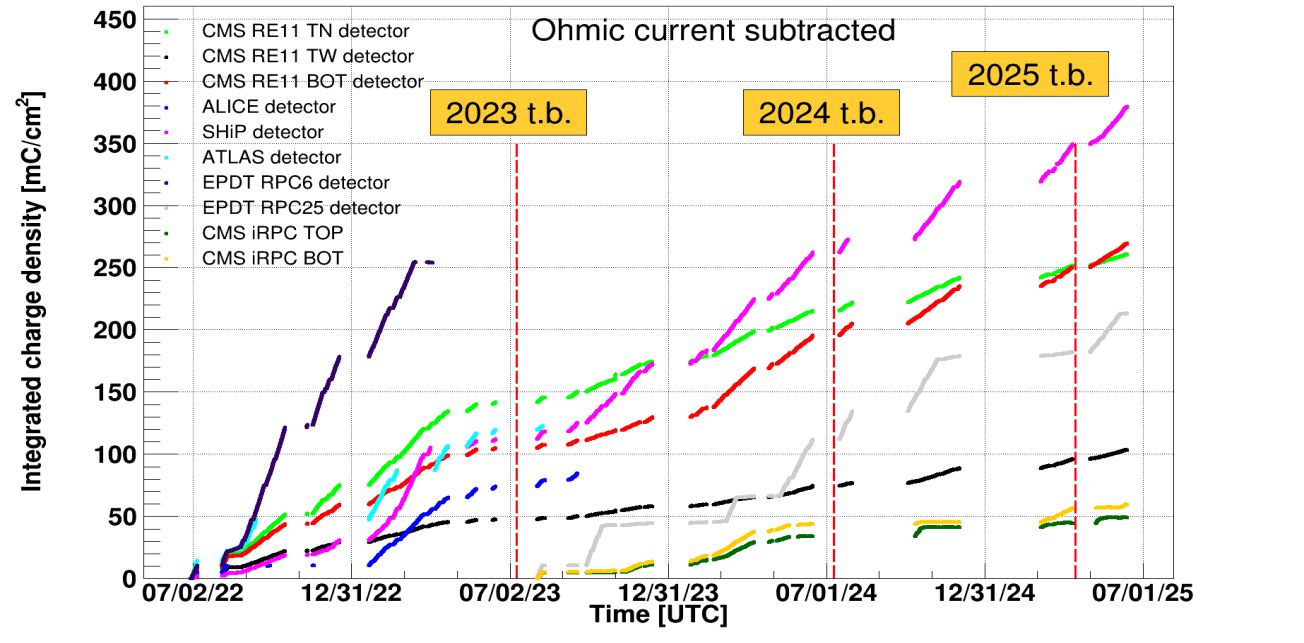}
    \caption{Integrated charge density, after subtraction of the Ohmic current, as a function of time for all the RPCs in the collaboration until July 2025. The vertical dashed lines indicate the 2023, 2024 and 2025 beam tests.}
    \label{fig:integrated-charge}
\end{figure}

The integrated charge is commonly used as a measure of the ageing an RPC has undergone. With the STD mixture, HF, that is produced at a rate roughly proportional to the current flowing through the gas, can damage the inner electrode surfaces, while in HFO-based mixtures the production of TFA may have a similar effect~\cite{Young2010}. Figure~\ref{fig:integrated-charge} shows the charge integrated by all the RPCs, after subtraction of the Ohmic current, from the start of the irradiation campaign until July 2025. The LHCb/SHiP RPC reached $\approx$380~mC/cm$^2$ of charge integrated, followed by the CMS RE11 BOT and TN gaps ($\approx$270 and $\approx$260~mC/cm$^2$) and the EP-DT RPC25 chamber ($\approx$210~mC/cm$^2$). The CMS RE11 TW gap integrated only $\approx$100~mC/cm$^2$, due to its lower current at the irradiation voltage. For comparison, the integrated charge targets are $\approx$100~mC/cm$^2$ for ALICE and $\approx$1~C/cm$^2$ for CMS, including a safety factor of 3.

\begin{figure}[htbp]
    \centering
    \includegraphics[width=0.45\textwidth]{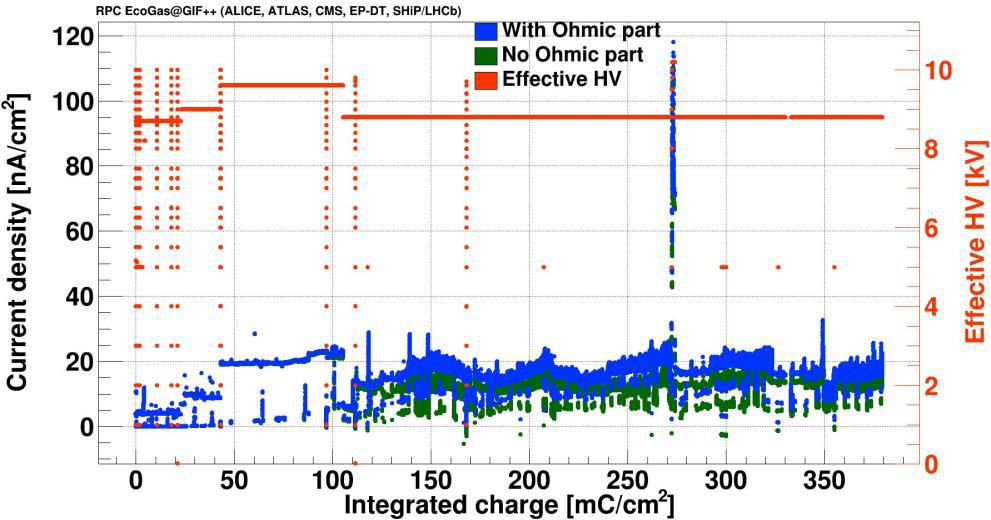}
    \hfill
    \includegraphics[width=0.45\textwidth]{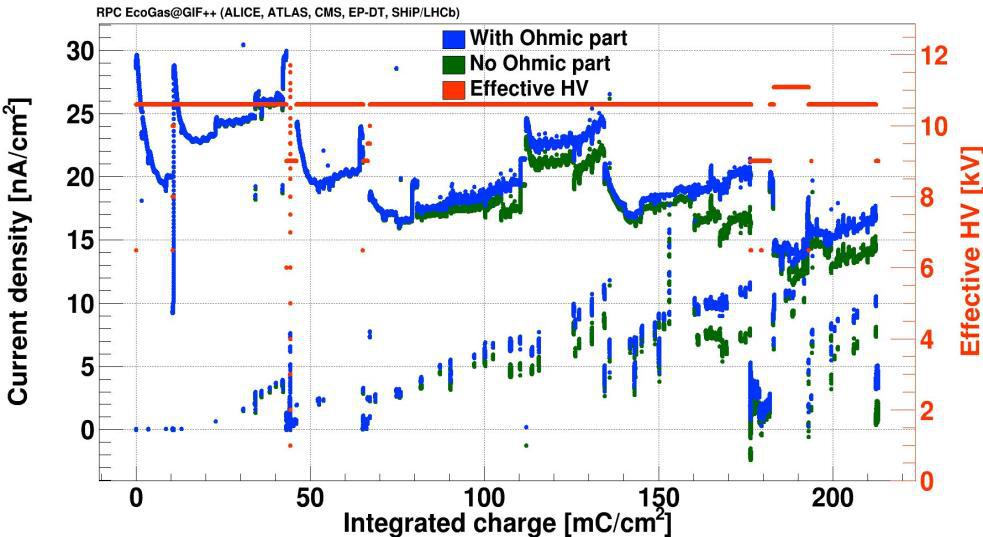}
    \caption{Current density with and without the Ohmic component (left axis) and
    $HV_\mathrm{eff}$ (right axis) versus integrated charge for the LHCb/SHiP (left)
    and EP-DT (right) detectors.}
    \label{fig:current-vs-charge}
\end{figure}

Figure~\ref{fig:current-vs-charge} shows both the total current density measured and the current density after subtraction of the ohmic contribution as a function of the integrated charge. For the LHCb/SHiP RPC (Left), In the first part of the campaign, the effective HV was increased in steps from $\approx$8.5 to $\approx$9.8~kV, and the total current under irradiation followed these steps, reaching $\approx$22--23~nA/cm$^2$ towards the 100~mC/cm$^2$ of integrated charge, with a negligible Ohmic contribution. At $\approx$105~mC/cm$^2$ the HV was reduced to 8.8~kV and kept constant thereafter since the total current drawn by this chamber was high. From this point, the current under irradiation fluctuates between $\approx$12 and 25~nA/cm$^2$, while the total current and current without ohmic contribution start to separate, indicating a non-negligible (a few nA/cm$^2$) but stable Ohmic dark current. The current spike observed at $\approx$270~mC/cm$^2$ occurred when the HFO bottle was close to being empty. By July 2025, the detector had integrated a charge of $\approx$295~mC/cm$^2$.
A different behavior is observed for the EP-DT detector, shown in Figure~\ref{fig:current-vs-charge} (right) and operated at 10.6~kV for almost the whole period. The current under irradiation first decreases, from $\approx$29 to $\approx$19~nA/cm$^2$, then slightly increases, and afterwards fluctuates between $\approx$15 and 25~nA/cm$^2$. By July 2025, the detector had integrated a charge of $\approx$165~mC/cm$^2$. The two detectors are representative of the behaviour observed for all the chambers.

\begin{figure}[htbp]
    \centering
    \includegraphics[width=0.5\textwidth]{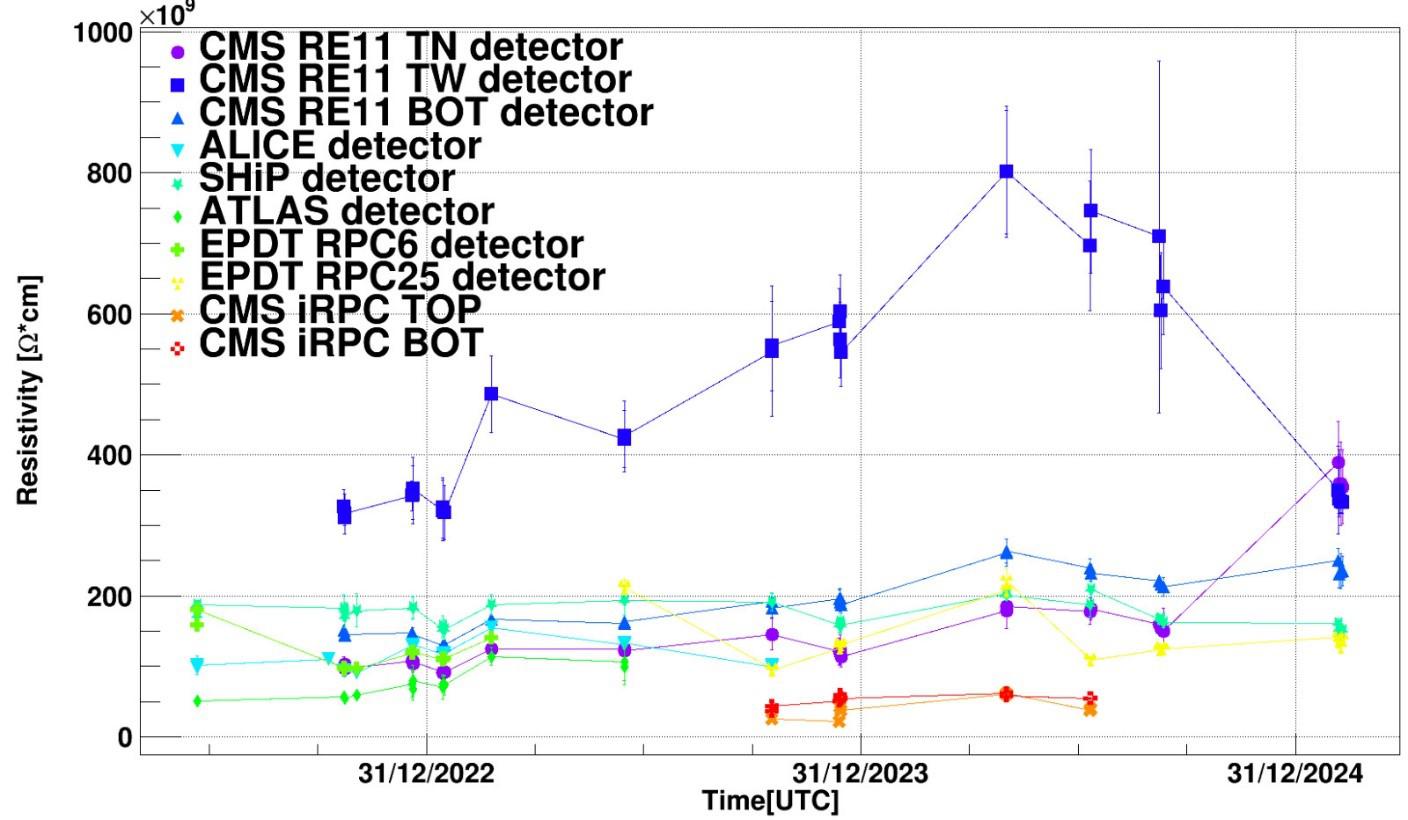}
    \caption{Evolution of the electrode resistivity over time for all the RPCs of the collaboration}
    \label{fig:Resitivity_all_detectors}
\end{figure}

The increase of the Ohmic current observed for both detectors in some integrated charge ranges could suggest a decrease of the electrode resistivity with time, since the two quantities are related through Ohm's law. To verify this assumption, the electrode resistivity of all detectors was periodically measured with the argon-based method described in~\cite{Aielli2006}. The results are reported in Figure~\ref{fig:Resitivity_all_detectors} as a function of time. No significant decrease in resistivity is observed, i.e. the value for the LHCb/SHiP RPC remains stable, while that of the EP-DT detector fluctuates without a clear trend. For the other detectors, the resistivity lies within $\approx$(50--250)$\times10^{9}$~$\Omega\cdot$cm, with a slow increase over time for some of them, except for the CMS RE11 TW gap, which shows higher values and larger uncertainties. Therefore, the increase of the Ohmic current cannot be attributed to a change in electrode resistivity, and its origin is currently under investigation. A possible contribution could come from damage to the inner surface of the electrodes, which is currently being examined for all the RPCs, in part also by means of chemical analyses.

\FloatBarrier
\section{Beam test results}
\label{sec:beamtest}

\subsection{Performance without irradiation}

The efficiency of each detector is measured as a function of the effective high voltage
and fitted using a sigmoid function:

\begin{equation}
\label{eq:sigmoid}
\mathcal{E}(HV^{}_{\rm eff}) = \frac{\mathcal{E}_{\rm max}}{1+e^{-\beta\left(HV^{}_{\rm eff}-HV_{50}\right)}},
\end{equation}

where $\mathcal{E}_{\rm max}$ denotes the plateau efficiency,  the voltage $HV_{50}$ at which the efficiency reaches 50\% of $\mathcal{E}_{\rm max}$, and the slope parameter $\beta$ describing the steepness of the turn-on of the curve. The working point (WP) is defined as the voltage at which the efficiency reaches 95\% of $\mathcal{E}_{\rm max}$.

\begin{figure}[H]
    \centering
    \includegraphics[width=.5\textwidth]{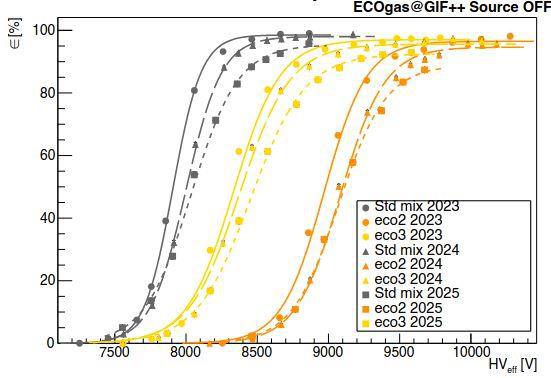}
    \caption{Efficiency curves as a function of $HV_{\mathrm{eff}}$ for LHCb/SHiP chamber.}
    \label{fig:eff-noirr}
\end{figure}

Figure~\ref{fig:eff-noirr} compares the source-off efficiency curves of the LHCb/SHiP chamber measured between 2023 and 2025 with the STD, ECO2, and ECO3 mixtures. After two years of irradiation, the curves exhibit a slight reduction in plateau efficiency and a shift towards higher effective voltages. For all measurement periods, the working point follows the ordering $\mathrm{WP}(\mathrm{ECO2}) > \mathrm{WP}(\mathrm{ECO3}) > \mathrm{WP}(\mathrm{STD})$, indicating that both eco-friendly mixtures require higher operating voltages than the standard mixture, with ECO2, having the larger HFO-fraction, showing the largest increase.

\begin{figure}[H]
    \centering
    \includegraphics[width=0.4\textwidth]{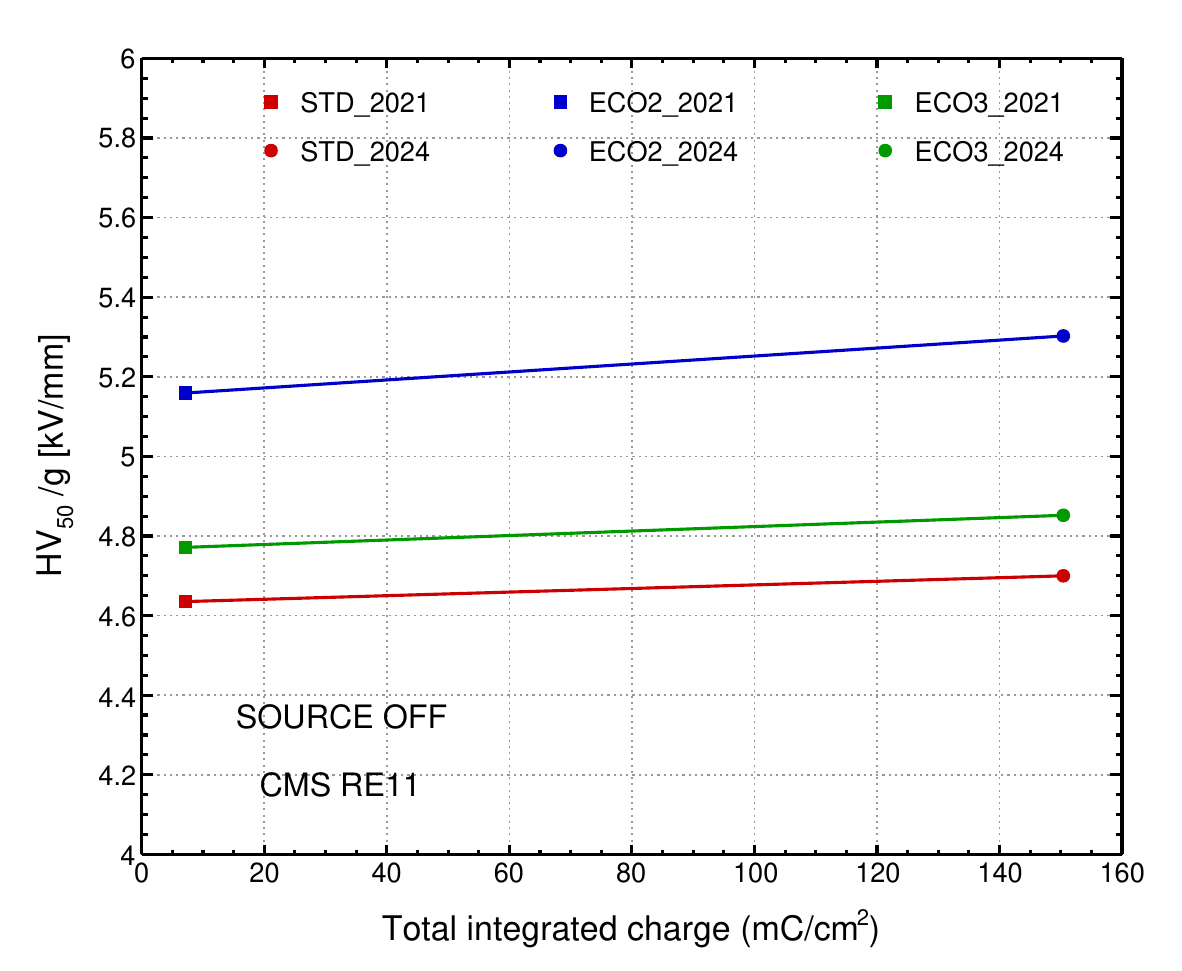}
    \hfill
    \includegraphics[width=0.4\textwidth]{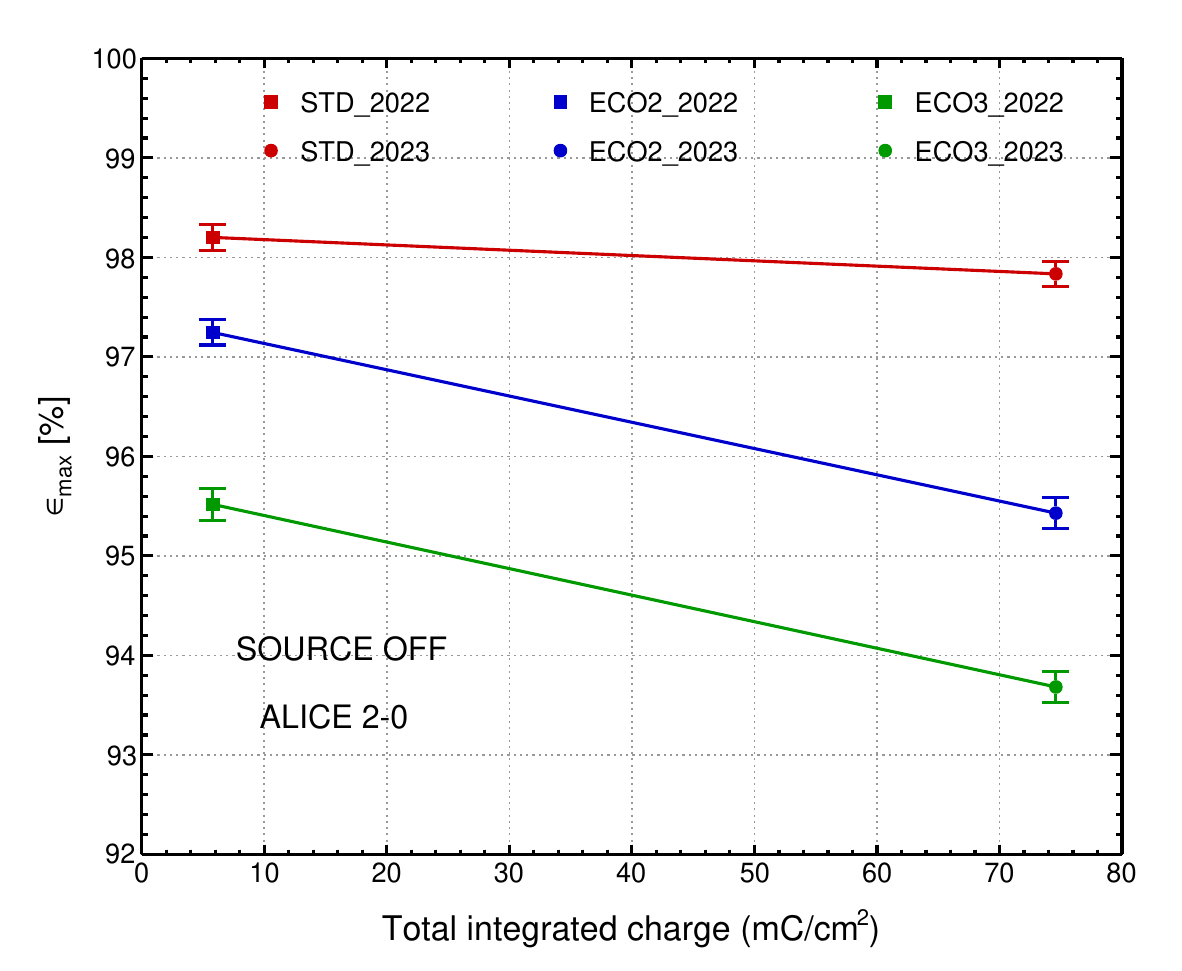}
    \caption{Evolution of $HV_{50}/g$ (left) and maximum efficiency
    $\varepsilon_{\max}$ (right) as a function of integrated charge.}
    \label{fig:hv50-emax-vs-charge_sourceoff}
\end{figure}

In order to assess the detector behavior, the plateau efficiency and $HV_{50}/g$ as a function of integrated charge before and after irradiation campaign are shown in Figure~\ref{fig:hv50-emax-vs-charge_sourceoff}. The value of $HV_{50}/g$ increases slightly after 3 years of irradiation at source off, with the largest shift observed for ECO2 mixture. For ALICE 2-0 detector, the plateau efficiency decreases marginally, with a more pronounced reduction for ECO2 and ECO3 gas mixtures than for STD mixtures. 
These source-off measurements reveal slight changes in detector response after ageing campaign.

\begin{figure}[H]
    \centering
    \includegraphics[width=0.32\textwidth]{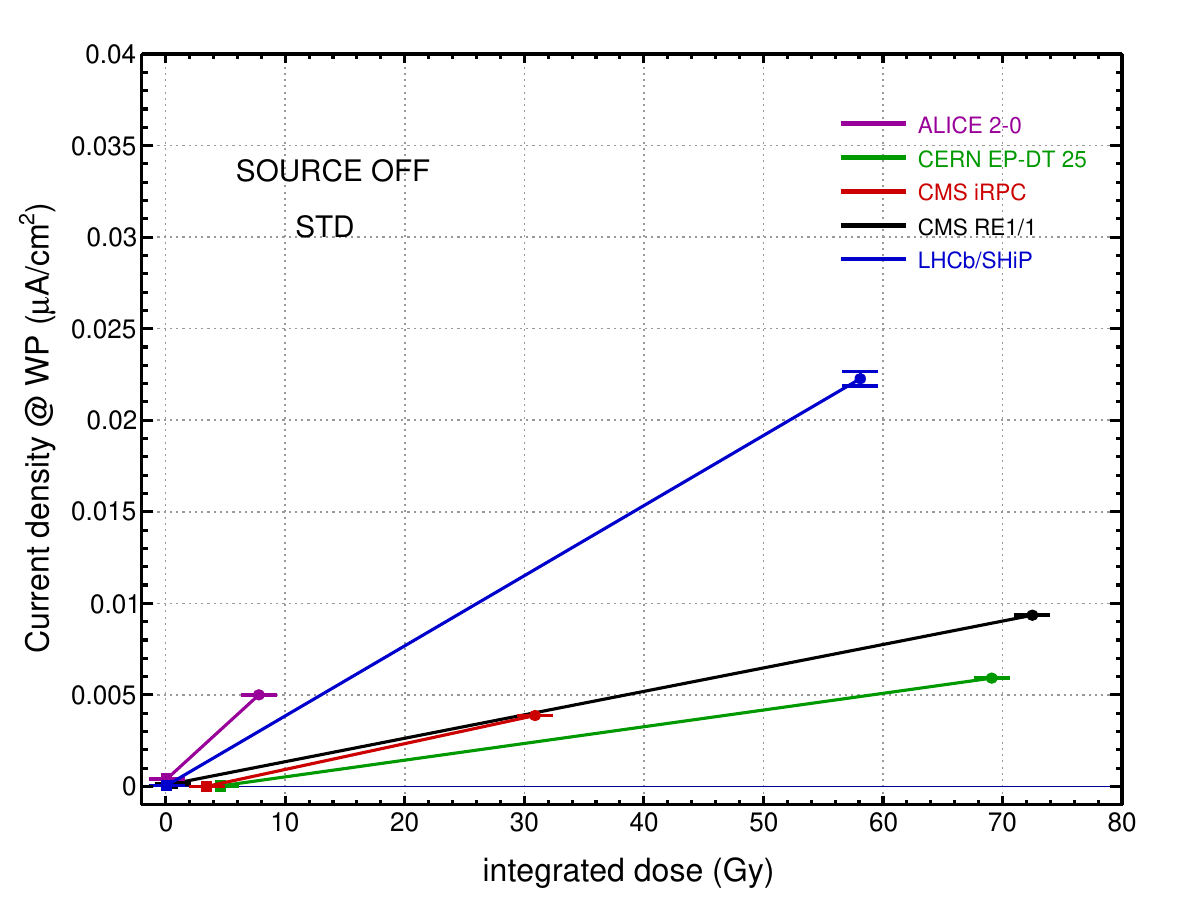}
    \hfill
    \includegraphics[width=0.32\textwidth]{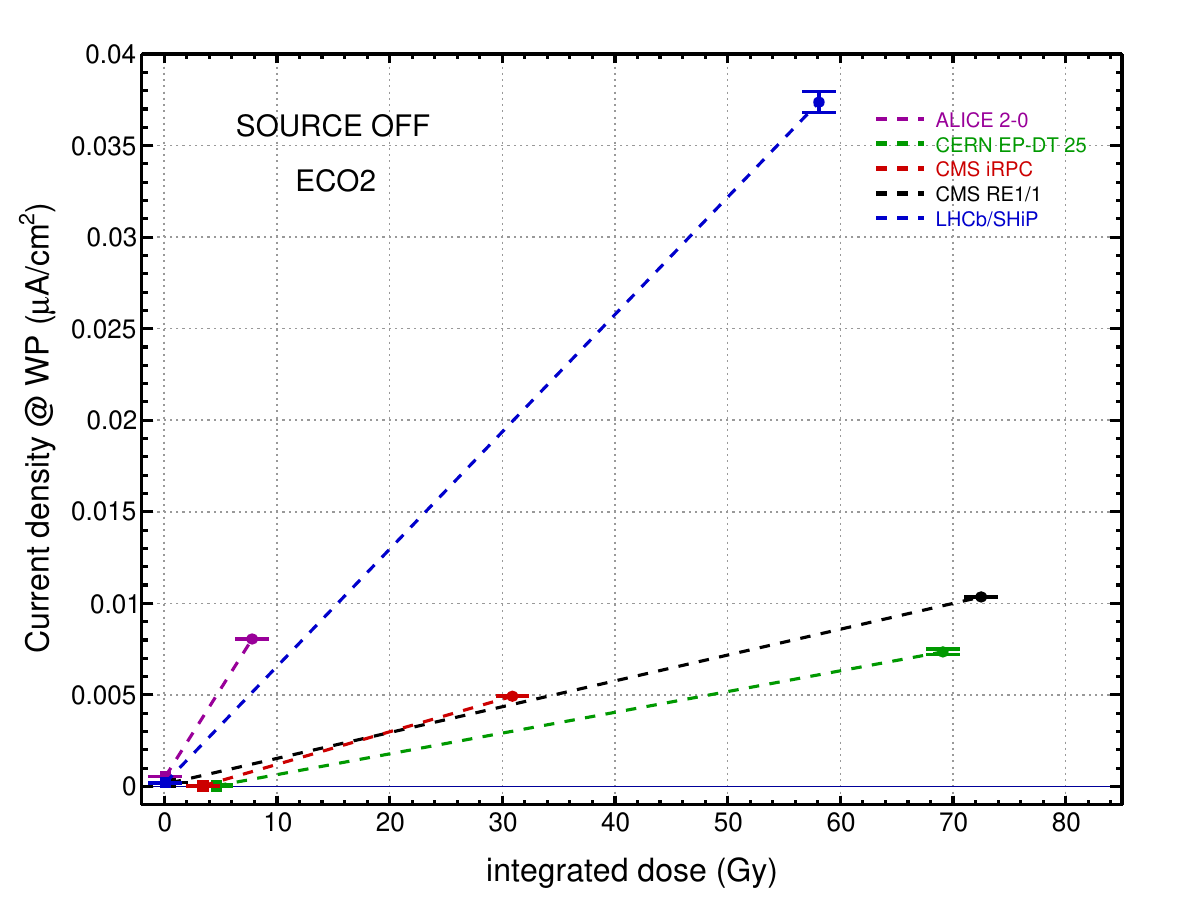}
    \hfill
    \includegraphics[width=0.32\textwidth]{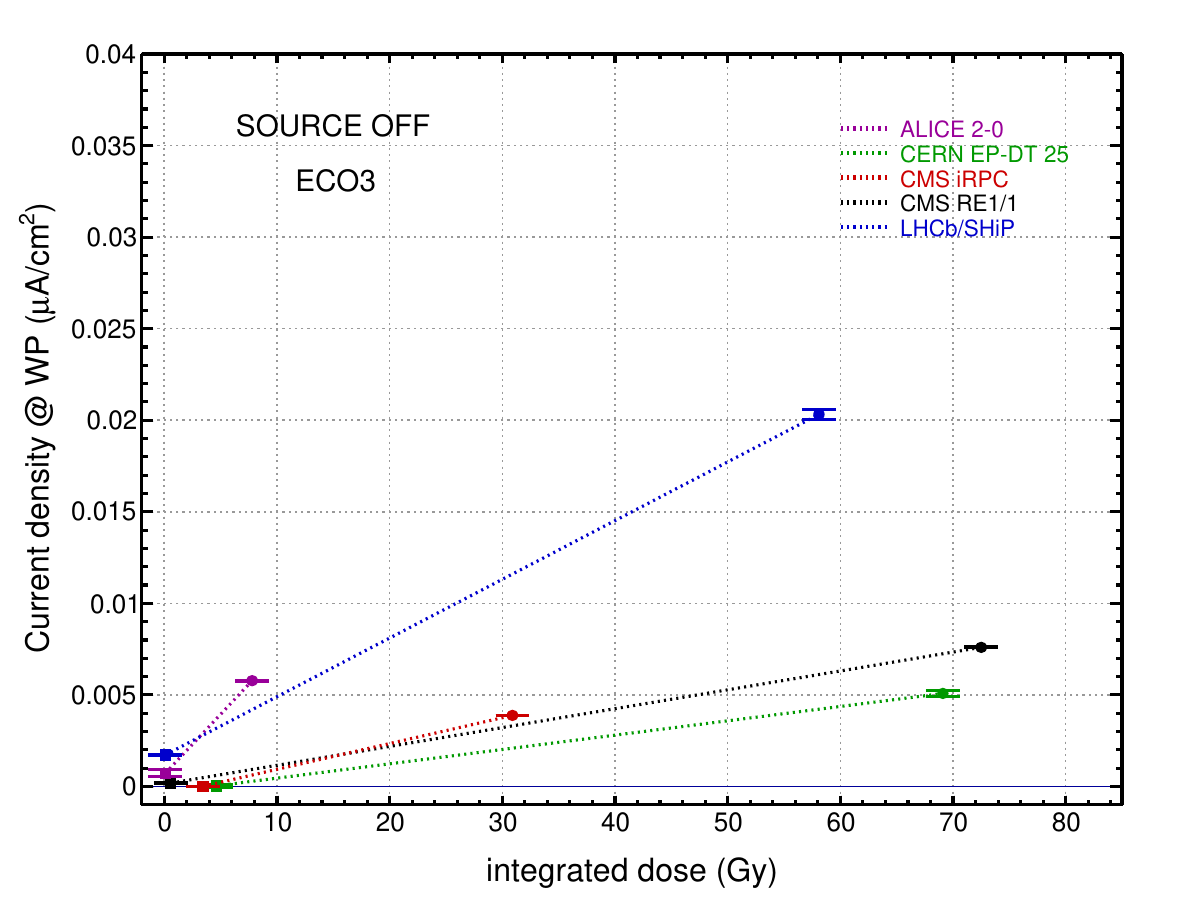}
    \caption{Current density at the working point (WP) as a function of integrated
dose for all chambers. The left, middle, and right
panels show the results for the STD, ECO2, and ECO3 gas mixtures, respectively.}
    \label{fig:current-source-off}
\end{figure}

To quantify the evolution of the chamber current without irradiation,
Figure~\ref{fig:current-source-off} presents the current density measured at
the working point as a function of integrated dose for all chambers and the
three gas mixtures studied. A clear increase in current density for increasing integrated dose is observed
for most chambers under test. In general, this increase is more pronounced
for chambers showing larger shifts of the efficiency curves towards higher
values of effective high voltage.

\FloatBarrier
\subsection{Performance during irradiation}

\begin{figure}[H]
    \centering
    \includegraphics[width=0.5\textwidth]{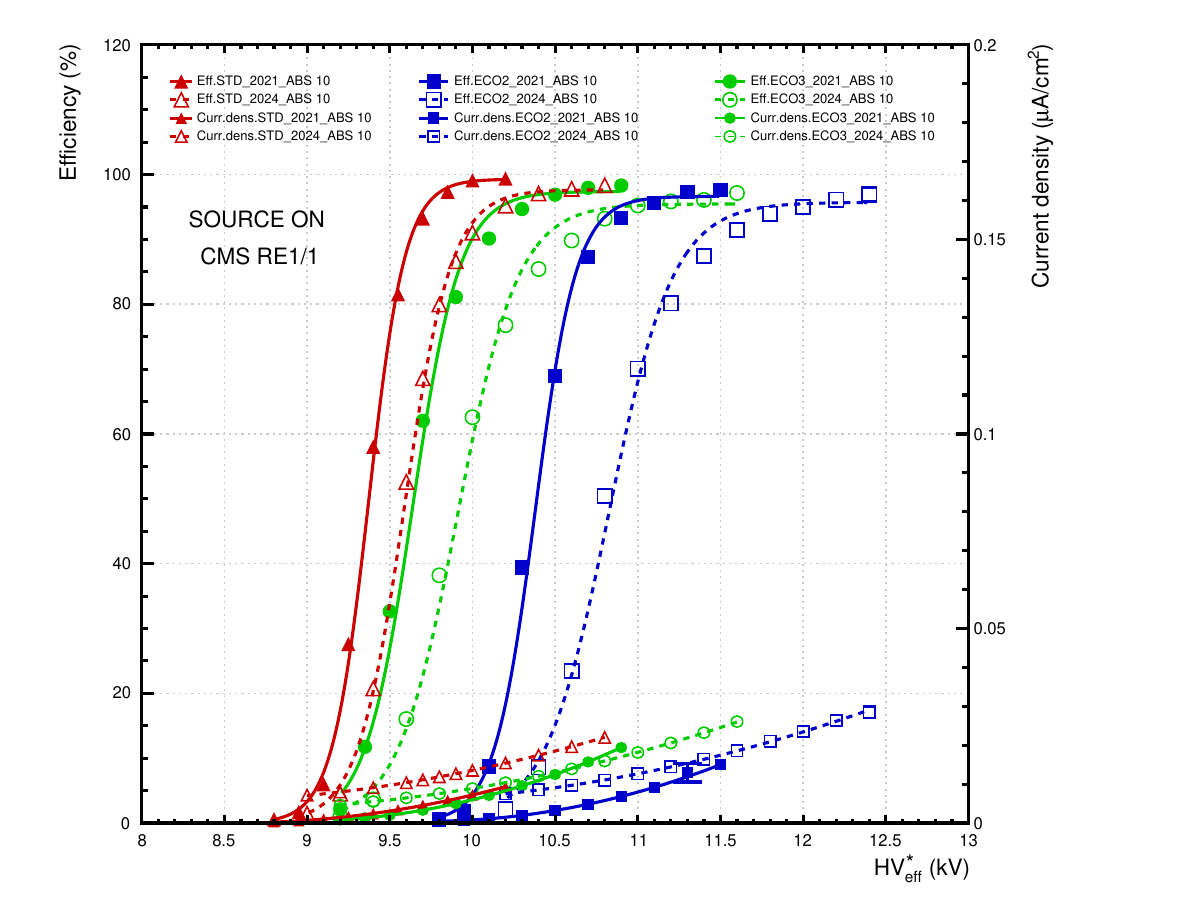}
    \caption{Efficiency and current density as a function of effective high voltage
    for the CMS RE1/1 detector.}
    \label{fig:eff-irradiation}
\end{figure}

Studies of the evolution of the detector efficiency and current density under high-rate conditions were also performed. Figure~\ref{fig:eff-irradiation} compares the efficiency and current density of the CMS RE1/1 chamber measured in 2021 (before irradiation) and 2024 (after irradiation) for the STD, ECO2, and ECO3 gas mixtures. Between the two measurements, the chamber accumulated an integrated dose of $\approx$72.5 Gy, whereas the dose before the start of the irradiation campaign was negligible.  Both measurements were performed with the $^{137}$Cs  source operated at a filter setting of $\mathrm{ABS}=10$, corresponding to a background $\gamma$ rate of approximately 250 -- 350~Hz/cm$^2$. After irradiation, the shift towards higher effective voltages and the slight decrease in the steepness of the efficiency curves are both more pronounced for ECO2 and ECO3 gas mixtures than for STD. An increase in current density is also visible in the later measurements.

\begin{figure}[H]
    \centering
    \includegraphics[width=0.4\textwidth]{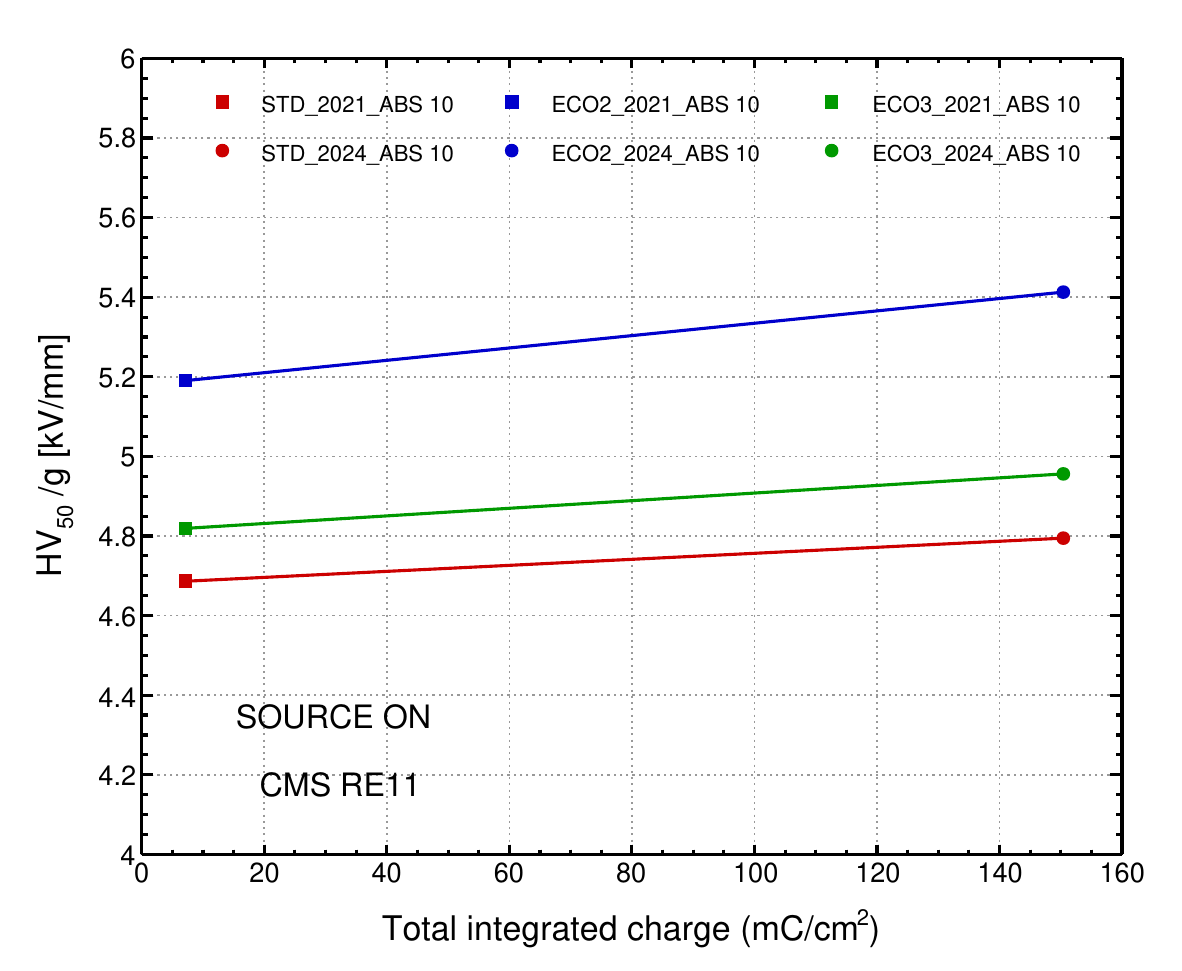}
    \hfill
    \includegraphics[width=0.4\textwidth]{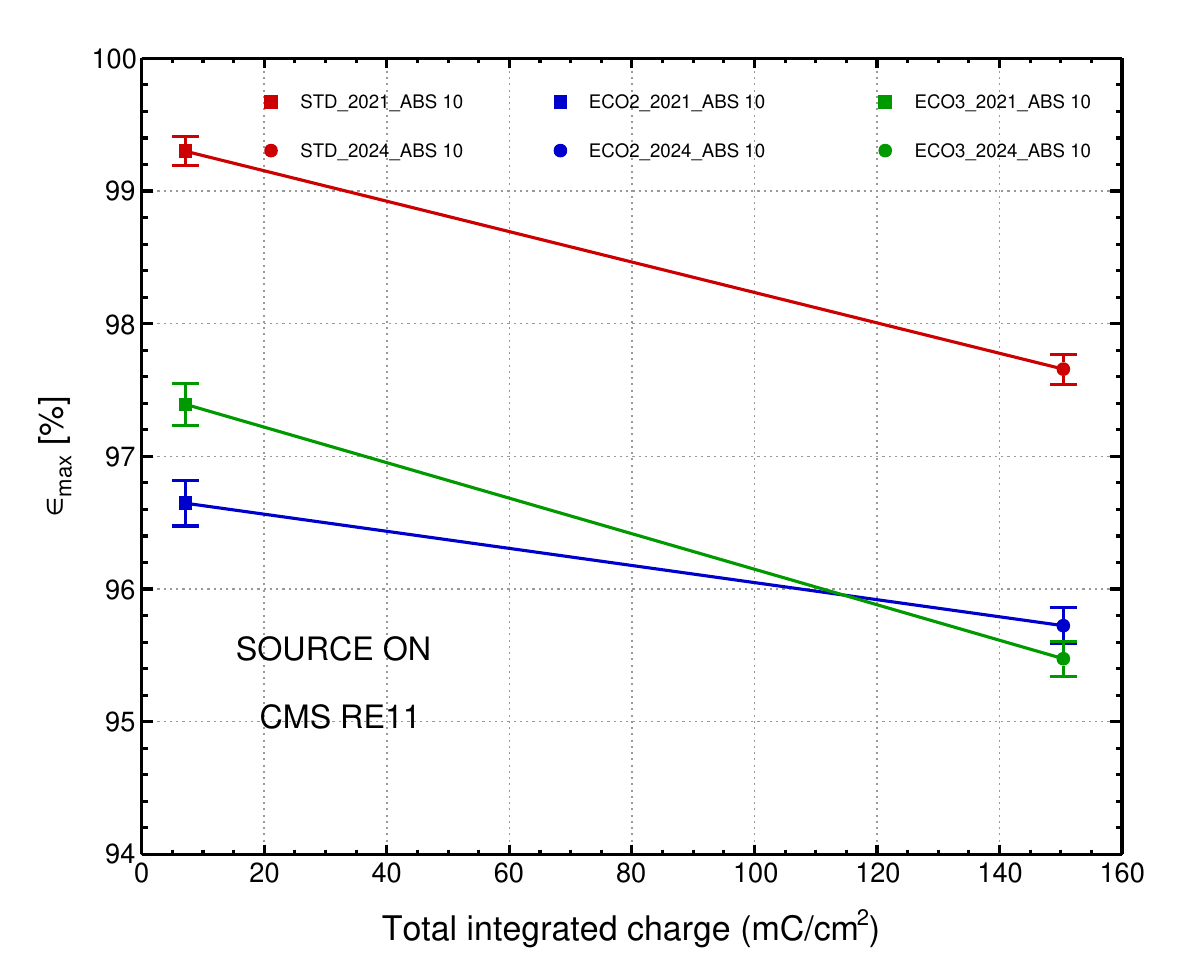}
    \caption{Evolution of $HV_{50}/g$ (left) and plateau efficiency
    $\varepsilon_{\max}$ (right) as a function of integrated charge for the
    CMS RE1/1 detector, measured with the source on at $\mathrm{ABS}=10$
    for the STD, ECO2, and ECO3 mixtures.}
    \label{fig:hv50-emax-vs-charge_sourceon}
\end{figure}

To quantify these changes, Figure~\ref{fig:hv50-emax-vs-charge_sourceon} shows the $HV_{50}/g$, and the plateau efficiency, $\varepsilon_{\max}$, as a function of integrated charge for the same chamber and filter setting. As observed in the measurements without irradiation, $HV_{50}/g$ increases for all three mixtures, with the largest shift observed for ECO2. The plateau efficiency decreases slightly, with the largest reduction observed for ECO3. These measurements indicate that high efficiency is maintained under irradiation, although higher operating voltages are required after ageing.

\begin{figure}[H]
    \centering
    \includegraphics[width=0.35\textwidth]{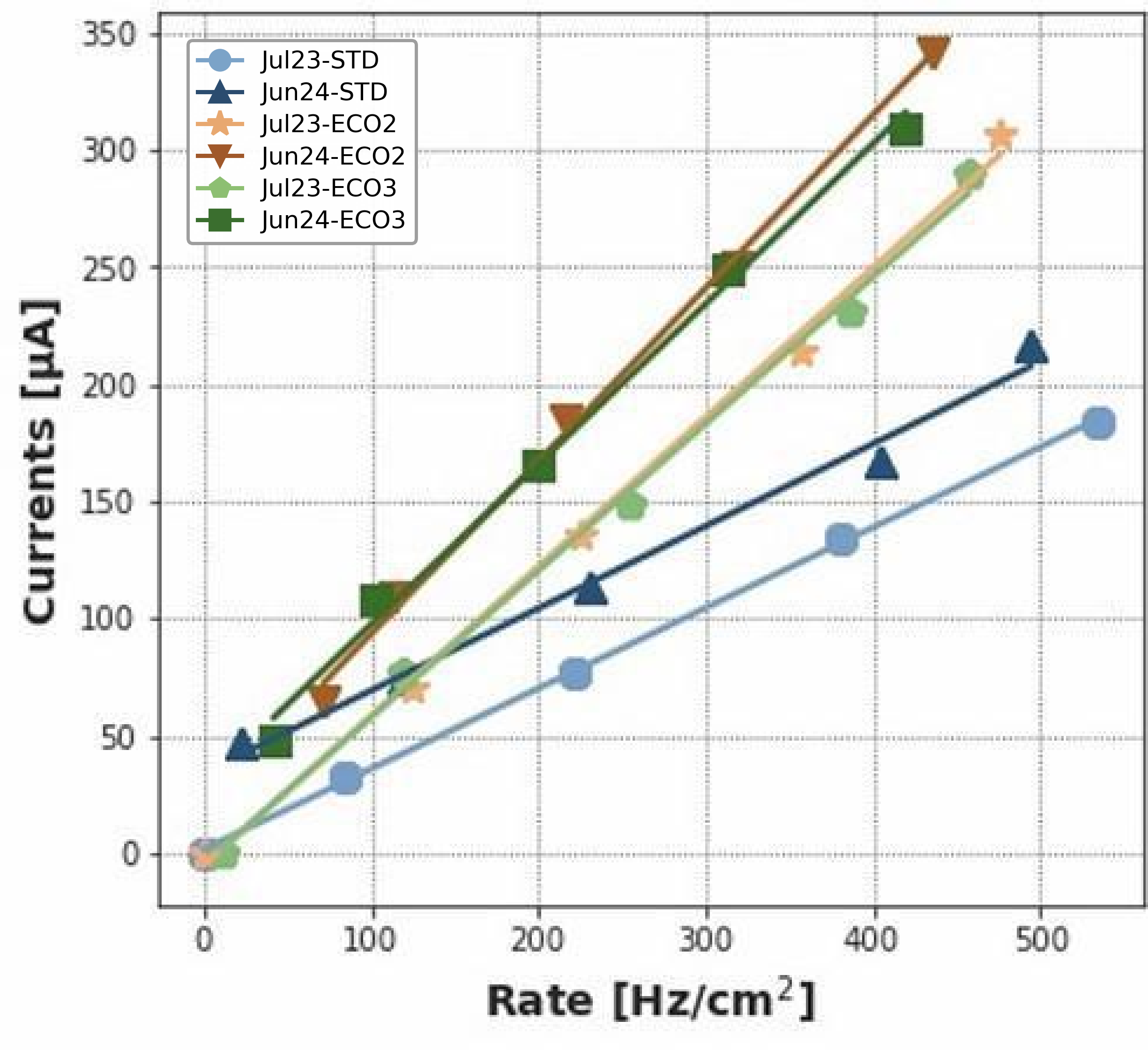}
    \hfill
    \includegraphics[width=0.35\textwidth]{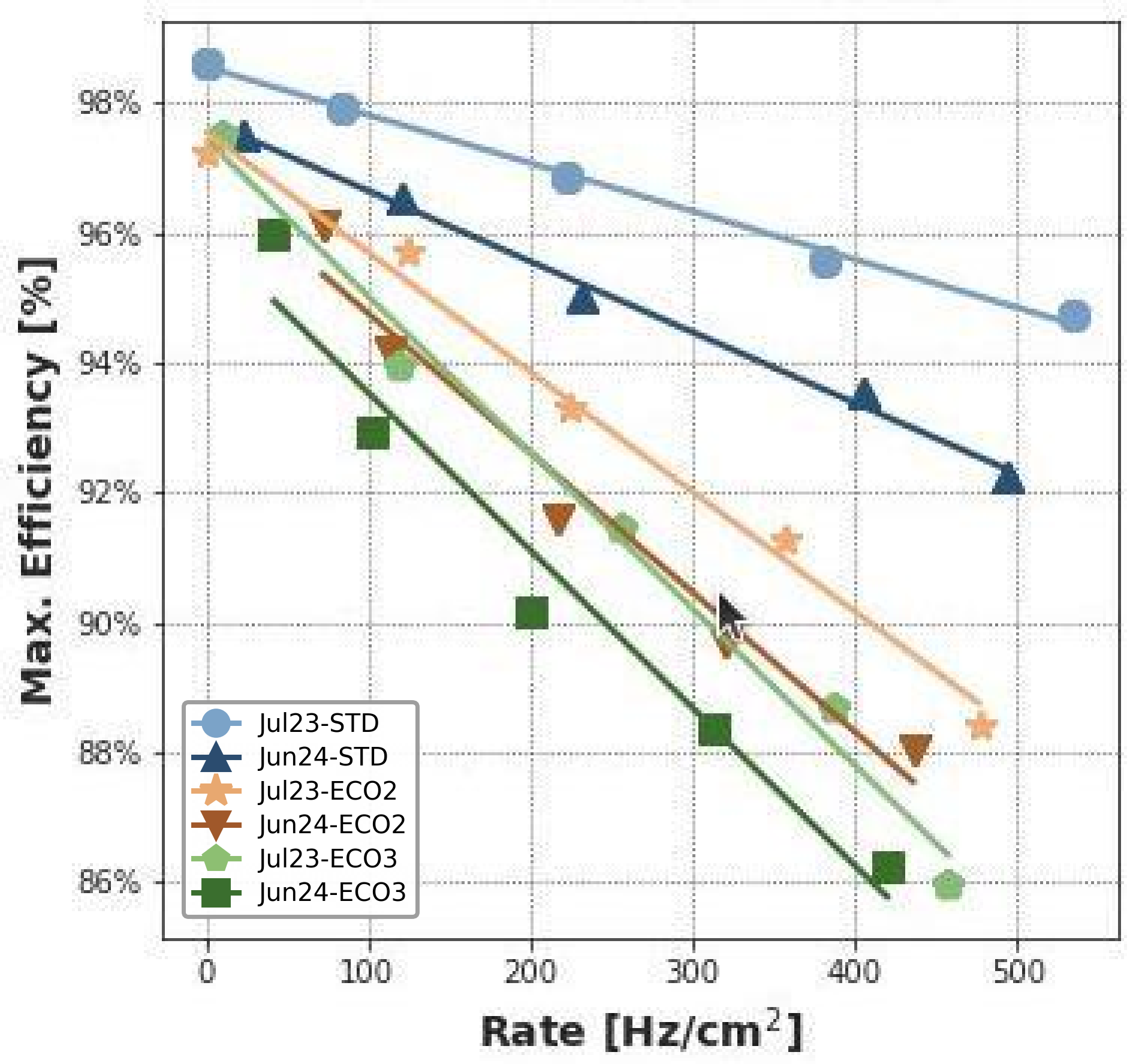}
    \caption{Current (left) and maximum efficiency (right) as a function of
    $\gamma$ rate for the EP-DT detector.}
    \label{fig:epdt-rate-dependence}
\end{figure}

Complementary measurements at different irradiation rates are shown in Figure~\ref{fig:epdt-rate-dependence} for the EP-DT detector, a single-gap RPC with a 2~mm gas gap, after approximately $115~\mathrm{mC/cm^{2}}$ of integrated charge. The current increases with background rate for all three gas mixtures, with higher currents observed in 2024 than in 2023 at comparable rates. Correspondingly, the maximum efficiency decreases with increasing $\gamma$ rate after irradiation.

\FloatBarrier
\section{Summary and future perspective}
\label{sec:summary}
The RPC ECOGas@GIF++ Collaboration has performed a long-term ageing study at CERN GIF++ on RPCs with different gap thicknesses and readout electronics, irradiated with HFO-1234ze/CO$_2$-based mixtures since July 2022.

Beam tests show that high chamber efficiencies are maintained after irradiation, with total integrated doses between about 8 and 73 Gy depending on the detector, for the STD (95.2\% C$_2$H$_2$F$_4$, 4.5\% i-C$_4$H$_{10}$, 0.3\% SF$_6$), ECO2 (35\% HFO-1234ze, 60\% CO$_2$, 4\% i-C$_4$H$_{10}$, 1\% SF$_6$) and ECO3 (25\% HFO-1234ze, 69\% CO$_2$, 5\% i-C$_4$H$_{10}$, 1\% SF$_6$) mixtures. The main effects are a shift of the working point towards higher effective voltages and a reduced plateau efficiency, accompanied by an increase in the chamber current at the working-point voltage, whose origin is still under investigation. No severe degradation attributable to the HFO-based mixtures has been observed, supporting their use as eco-friendly alternatives for RPC operation.

Future work will correlate the efficiency parameters with integrated charge and absorbed dose, investigate the dark current oscillations and include post-irradiation studies of the electrode materials. In parallel, R1233zd (GWP $\approx 1$) is being studied as an SF$_6$ replacement; several RPCs were operated with such mixtures in a recent GIF++ beam test campaign, and their performance is currently being evaluated. 

Overall, the studies reported here are essential to qualify eco-friendly gas mixtures for long-term operation of RPCs in future high-luminosity experiments.

\acknowledgments
This work was supported in part by the Research Council of the VUB in Belgium (project OZR4149).


\bibliographystyle{JHEP}
{\small
\bibliography{biblio}
}
\end{document}